# Information-theoretic point symmetry classifications/quantifications of an electron diffraction spot pattern from a crystal with strong translational pseudosymmetry

Peter Moeck and Lukas von Koch

*Abstract*—The recently developed information-theoretic approach to crystallographic symmetry classifications and quantifications in two dimensions (2D) from digital transmission electron and scanning probe microscope images is adapted for the analysis of an experimental selected-area transmission electron diffraction spot pattern from $Ba_3Nb_{16}O_{23}$. The extracted lattice parameters of this crystal are within experimental error bars consistent with a metric tensor that suggests the presence of hexagonal translation symmetry. The point symmetry of the combined low, medium, and high resolution spots is, however, no higher than *2mm*. The likelihood of this electron diffraction pattern belonging to a rectangular-centered crystal rather than a hexagonal crystal is quantified on the basis of its information-theoretic point group symmetry classifications. Presumably due to a slight misorientation away from the exact [001] zone axis combined with the curvature of the Ewald sphere and a real structure that includes intergrowth of quadruple NbO and triple $BaNbO_3$ "blocks", the group of highest resolution spots (with interplanar spacings between 1.25 to 0.85 Å), feature point symmetry *.m.* only. The crystallographic $R_{sym}$ values of traditional classifications into the point groups that are compatible with the experimentally obtained primitive lattice parameters are provided for comparison purposes. As it is common practice in diffraction based crystallography, point symmetry classification/quantification results for the group of highest resolution spots are provided separately from those for the combined low, medium, and high resolution spots.

## I. INTRODUCTION

Crystals with translational pseudosymmetries are not rare in nature [1-3]. Such pseudosymmetries complicate Bravais lattice type classifications of experimental data by methods such as powder X-ray diffraction, where point symmetry information on the crystallites in the powder is not explicitly revealed. It is often overlooked that the Bravais lattice types form a hierarchy [4] where the transition from a lower symmetric type to a higher symmetric type is marked by a metric specialization. The components of a metric tensor that has been derived from experimental measurements always feature error bars so that the lattice parameter values never obey the symmetry restrictions of the Bravais lattices types exactly [4]. An information-theoretic method has, therefore, been developed to assign Bravais lattice types to experimental data objectively [5]. This method is now complemented with an information-theoretic method for classifications and quantifications of 2D diffraction spot intensity data from transmission electron microscopes (TEMs) or Buerger precession X-ray diffraction cameras.

In this paper, objectivity is defined as being strictly based on the experimental data itself and reasonable assumptions about the generalized noise in the analysis [6]. Subjectivity, on the other hand, is here understood as using arbitrarily set thresholds on the deviation of a measurement value from a presumed symmetry dictated value. Such thresholds allow for the assignment of a certain Bravais lattice type by an investigator although the lattice parameters and components of the metric tensor feature error bars. Thresholds may be "hard coded" into an analysis program and the investigator may not readily be aware of their existence. Information-theoretic methods do not use such thresholds for the assignment of symmetries to experimental data and are, therefore, considered to be objective [6].

The new point symmetry classification/quantification method is conceptually similar to the information-theoretic method for plane symmetry group and projected Laue class classifications/quantifications from digital images that were recorded with transmission electron microscopes from crystals or scanning probe microscopes from crystal surfaces [7-9]. The major difference of the new method is that the experimental intensities of electron diffraction spots or essentially featureless (blank) non-overlapping disks are used for the calculations of sums of (normalized) squared residuals to suitably symmetrized counterpart intensities, whereas, it is the complex Fourier coefficients of the image intensity that provide the basis for the calculations in the classification/quantification method for digital images with respect to plane symmetry groups.

As the calculations of the sums of (normalized) squared residuals of the information-theoretic method are preformed in Fourier space, the projected Laue class [7,8] is obtained from the amplitude map of the discrete Fourier transform of the image intensity. These Laue classes are the 2D point symmetries that contain point group *2*, i.e. the projection of a 3D inversion center. Because the Fourier transform is centrosymmetric, the projected Laue class of a crystal with point symmetry *3m* is point group *6mm*. Analogously, the projected Laue class of a crystal with point symmetry *3* is point group *6*. Also, the amplitudes of the complex Fourier coefficients of a high-resolution TEM image from a crystal without an inversion center acquires projected Laue class *2* in 2D, just by virtue of calculating the discrete Fourier transform. There is, therefore, no sum of squared Fourier coefficient amplitude residuals for Laue class *2* in any crystallographic symmetry classification/quantification from digital images [7-9].

Making point symmetry classifications/quantifications from electron diffraction patterns rather than from the amplitude maps of discrete Fourier transforms of the corresponding images has three advantages. The first of these advantages is that there are typically many more Bragg reflections in an electron diffraction pattern than there are Fourier component amplitudes in the amplitude map of the discrete Fourier transform of a high-resolution image that was recorded from the same crystalline sample at the same location. This is because aberrations of the objective lens, defocus and alignment dependent attenuation functions as well as apertures, etc. are involved in the formation of an experi-

The first (and corresponding) author is with the Nano-Crystallography Group of the Department of Physics, Portland State University, Portland, OR, USA (phone: 503-725-4227; fax: 503-725-2815; e-mail: pmoeck@ pdx.edu). The second author volunteers in that research group and is currently a senior of Westside Christian High School, Portland, Oregon, USA (email: lukasvonkoch@gmail.com), but will soon be a freshman at the University of Pennsylvania.

mental high-resolution TEM image [10]. All of this restricts the effective Abbe resolution in Fourier space.

The second advantage is that a diffraction-mechanism induced two-fold rotation point is not automatically added to the projected point symmetry group of an experimental, 3D-information containing electron diffraction spot (or blank disk) patterns from merohedral crystals. There is, thus, a meaningful sum of squared residuals of the intensity of the diffraction spots for point group *2* for non-centrosymmetric crystals and one will often be able to distinguish point symmetries *3* and *6* from each other (as well as *2* and *4*, or *3m* and *6mm*, etc.) based on their geometric Akaike Information Criterion (G-AIC) values [7,8] and geometric Akaike weights [7,9].

Electron diffraction patterns are theoretically translation invariant if one were to obtain them from an ideal plan-parallel slab of an ideal crystal in an ideal TEM. This means that small random sample drifts do not affect a symmetry classification/quantification from an electron diffraction pattern in the same detrimental manner than that from a high-resolution TEM image. This is the third advantage of using electron diffraction spot (and blank disk) patterns for point symmetry classifications over Laue class classifications from high-resolution TEM images. Taking all of this into account, more accurate point symmetry classifications and quantifications can be made from electron diffraction spot patterns, such as the one shown in Fig. 1.

Finally, there is the widespread misconception that an experimentally observed translation symmetry would restrict the point and space group symmetry of a crystal. This is definitively not the case! By the Neumann-Minnigerode-Curie principle, it is the point symmetry of the crystal structure that restricts the symmetries of the macroscopic physics properties including the symmetry of the components of the metric tensor of a crystal (and the corresponding equality relationships of its lattice parameters).

This paper quantifies the evidence that the diffraction pattern in Fig. 1 originated from a crystal with a rectangular-centered Bravais lattice in spite of the extracted lattice parameters $a = 12.46 \pm 0.15$ Å, $b = 12.41 \pm 0.15$ Å, and $\gamma = 119.5 \pm 1°$. These parameters refer to an oblique Bravais lattice that is within error bars compatible with a hexagonal Bravais lattice. They were obtained with the well known electron crystallography program CRISP/ELD 2.1 [10] in its default setting. The second section of this paper provides in three subsections a brief account of the analysis of the electron diffraction pattern in Fig. 1. The third section mentions a few potential applications of the new method beyond obvious uses in electron crystallography [10] briefly. The paper ends with a summary and conclusions section.

## II. 2D POINT SYMMETRY ANALYSIS OF AN EXPERIMENTAL ELECTRON DIFFRACTION SPOT PATTERN

### A. The electron diffraction spot pattern and its analysis

Figure 1 presents the electron diffraction pattern that is classified and quantified in this paper with respect to its oriented 2D point symmetries. Note in passing that a few of the electron diffraction spots in the pattern in Fig. 1 are split, as marked by arrows. This indicates that the crystal features a "non-trivial" real structure, as indeed observed in [11] (in the form of intergrowth between quadruple NbO and triple $BaNbO_3$ "blocks"). In the context of this paper, the effects of this real structure are ignored because our focus is here to demonstrate the usefulness of information-theoretic point symmetry classifications and quantifications in the presence of generalized noise that includes a crystal's real structure.

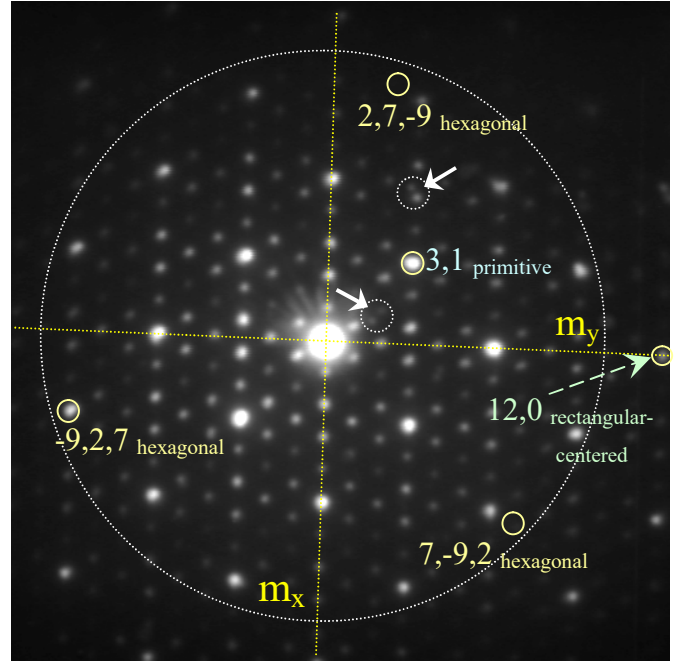

Figure 1. Electron diffraction spot pattern from $Ba_3Nb_{16}O_{23}$ in approximate [001] zone axis orientation. This pattern comes with the software CRISP/ELD and is discussed in [10,11]. Whereas the approximate vertical mirror line is $m_x$ (.*m*.), its approximate horizontal counterpart is $m_y$ (..*m*). The added circle signifies an Abbe resolution of 1.25 Å. Splitted electron diffraction spots that indicate a non-trivial real structure of the crytal are marked by arrows.

The diffraction spot intensities of the electron diffraction spots were extracted in the "shape integration" mode of CRISP/ELD and exported as *.hke files. Computer programs that were written by the second author of this paper took these *.hke files, amended them for missing spots, and calculated the normalized sums of squared intensity residuals, the G-AIC values, the likelihood that a particular geometric model with a certain point symmetry is the Kullback-Leibler (K-L) best model, the geometric Akaike weights [7-9], and the traditional (crystallographic) $R_{sym}$ [10] values for each of the nine possible geometric models of the spot intensity distribution in the pattern in Fig. 1. These geometric models are referred to by their oriented site/point symmetries in the tables below. The whole set of geometric models is compatible with the hexagonal Bravais lattice type.

### B. Results of the analysis of the pattern in Fig. 1

Tables IA and IB show results of the calculations of the above mentioned analysis programs out to an Abbe resolution of 1.25 Å. Results of the calculations of these programs for the region of the diffraction pattern with Abbe resolutions ranging from 1.25 to 0.85 Å are shown in Tables IIA and IIB.

The CRISP/ELD program failed to extract a few medium to low intensity reflections in the latter region, e.g. (12,0) in rectangular-centered indexing at 1 Å on mirror line $m_y$ in Fig. 1. In standard crystallographic analyses, the traditional $R_{sym}$ values of the highest resolution reflections are typically quoted separately and are often much larger that their counterparts for the combined low, medium, and high resolution reflections.

We follow this practice with Tables IB and IIB and contrast the $R_{sym}$ values with the geometric Akaike weights there.

TABLE IA. RESULTS FOR LOW, MEDIUM, AND HIGH RESOLUTION REFLECTIONS OUT TO 1.25 Å, 256 SPOTS

| Site symmetry of geometric model | G-AIC value (bit) | Likelihood to be K-L best model |
|---|---|---|
| 2 | 2.133316644 | 0.754468 |
| .m. | 1.756044468 | 0.911097 |
| ..m | 1.765226226 | 0.906923 |
| **2mm** | **1.56983177** | **1.0** |
| 3 | 8.658509194 | 0.0288877 |
| 3m1 | 8.336693093 | 0.0339308 |
| 31m | 8.341743807 | 0.0338453 |
| 6 | 8.431851947 | 0.0323542 |
| 6mm | 8.225889826 | 0.0358637 |

TABLE IB. RESULTS FOR LOW, MEDIUM, AND HIGH RESOLUTION REFLECTIONS OUT TO 1.25 Å, 256 SPOTS, CONTINUED

| Site symmetry of geometric model | Geometric Akaike weight (%) | Traditional $R_{sym}$ (%) |
|---|---|---|
| 2 | 16.7964208 | 15.1 |
| .m. | 20.28338148 | 11.7 |
| ..m | 20.19047634 | 11.3 |
| **2mm** | **22.26260251** | **15.9** |
| 3 | 0.643115812 | 52.2 |
| 3m1 | 0.7553890169 | 52.2 |
| 31m | 0.7534837966 | 52.2 |
| 6 | 0.7202896652 | 52.3 |
| 6mm | 0.7984197839 | 52.2 |
| | Sum = 100% | Sum is meaningless |

TABLE IIA. RESULTS FOR HIGHEST RESOLUTION REFLECTIONS BEYOND 1.25 Å, 174 SPOTS, MOST OF WHICH WITH VERY LOW INTENSITY

| Site symmetry of geometric model | G-AIC value (bit) | Likelihood to be K-L best model |
|---|---|---|
| 2 | 1.09439 | 0.695011 |
| **.m.** | **0.36673** | **1** |
| ..m | 0.998738 | 0.729057 |
| 2mm | 0.98106 | 0.735529 |
| 3 | 0.46228 | 0.953348 |
| 3m1 | 0.394519 | 0.986202 |
| 31m | 0.409891 | 0.978651 |
| 6 | 0.435334 | 0.96628 |
| 6mm | 0.388729 | 0.989061 |

TABLE IIB. RESULTS FOR HIGHEST RESOLUTION REFLECTIONS BEYOND 1.25 Å, 174 SPOTS, MOST OF WHICH WITH VERY LOW INTENSITY, CONTINUED

| Site symmetry of geometric model | Geometric Akaike weight (%) | Traditional $R_{sym}$ (%) |
|---|---|---|
| 2 | 8.6518 | 59.4 |
| **.m.** | **12.4484** | **29.3** |
| ..m | 9.07561 | 49.7 |
| 2mm | 9.15619 | 61.5 |
| 3 | 11.8677 | 68.3 |
| 3m1 | 12.2767 | 72.8 |
| 31m | 12.1827 | 70.1 |
| 6 | 12.0287 | 73.4 |
| 6mm | 12.3123 | 74.7 |
| | Sum = 100% | Sum is meaningless |

## C. Discussion of the analysis results of the pattern in Fig. 1

The G-AIC values in Tables IA and IIA were obtained for a total of nine geometric models of the point symmetry of the intensity of the spots in the diffraction pattern of Fig. 1. They are (normalized) sums of squared intensity residuals that are corrected for their specific geometric model selection bias [7-9]. The lower symmetric models of the intensities of the diffraction pattern spots (with two or three symmetry operations in their point group) always have the lower sums of squared residuals, simply because they have more degrees of freedom as they are less constraint by symmetries. The lower residual sums of the lowest symmetric models are in a statistically sound manner [6] counterbalanced by higher geometric model selection penalties.

Out of the nine geometric models for the diffraction spot intensities in Tables IA and IIA, the model with the lowest G-AIC value is declared the K-L best model because it minimizes the Kullback-Leibler divergence and is maximally supported by the experimental data. For the central region of the diffraction pattern in Fig. 1 out to Abbe resolution 1.25 Å, as marked by a circle, the K-L best model is the one that features point symmetry group *2mm*, see Table IA. The central region of the diffraction pattern is, accordingly, classified as belonging to *2mm* with an average confidence level of 38.83% over its maximal subgroups.

The three individual point symmetries that contribute to *2mm* are visually broken in that region of the diffraction pattern, but the individual breakings are in a manner that in aggregate the higher symmetric geometric model emerges as the best representation of the symmetry information. Low G-AIC values and high likelihoods are in Table IA also listed for the models with point groups *2*, *.m.*, and *..m* as these groups are the three maximal subgroups of *2mm*.

Whereas the likelihood of being the K-L best model is for the model of the experimental data that features point symmetry group *2mm* unity (as it simply is that model), the geometric models that feature its three maximal subgroups feature likelihoods smaller than unity. All of the geometric models that feature the other point groups that are compatible with hexagonal translation symmetry, i.e. *3*, *3m1*, *31m*, *6* and *6mm*, in Table IA have very small likelihoods of being the K-L best model.

Likelihoods enable the calculation of evidence ratios for the preference of one geometric model over another [7,9]. The quantification of the evidence in favor for the diffraction pattern in Fig. 1 belonging to a crystal with a rectangular-centered Bravais lattice is obtained from the ratio of the sum of the likelihoods of the models with point groups *.m.*, *..m*, and *2mm* to the sum of the likelihoods of the models with point groups *3*, *3m1*, *31m*, *6*, and *6mm*. This ratio is for the central region in the diffraction pattern in Fig. 1 approximately 17.1 to 1, i.e. rather overwhelming.

For the outer region of the diffraction pattern in Fig. 1, i.e. the spots beyond Abbe resolution 1.25 Å, the K-L best model for the experimental data is the one that features point symmetry group *.m.*, see Table IIA. In the figure itself, this mirror line is referred to as $m_x$ and approximately relates the left half of the pattern to its right half. The mirror line $m_y$ in Fig. 1, on the other hand, relates approximately the upper and lower half of the diffraction pattern. Clearly, there are many more spots visible in the lower half of the outer region of the diffraction patter than in the upper half. The $m_y$ mirror line must, therefore, be more severely broken than the $m_x$ mirror line already by visual inspection. This is in agreement with the entries for .m. and ..m in Table IIA.

The geometric Akaike weights in Tables IB and IIB are the probabilities that a certain geometric model is the K-L best model. For the whole set of models, the sum of these probabilities is per definition 100%. Sums of $R_{sym}$ values are, on the other hand, meaningless. Note that high values of geometric Akaike weights are "somehow associated" with low values for the traditional $R_{sym}$ in Table IB, but not in Table IIB. The lack of "association" in the latter table is obviously due to many spots being absent in the outer region of the diffraction pattern in Fig. 1.

On the basis of the $R_{sym}$ values, one would need to resort to arbitrary set thresholds (and the associated leaps of faith) to conclude that the central region of the diffraction pattern features point symmetry *2mm*. Similarly, one is likely to overlook that the point symmetries *2* and ..m are so severely broken in the outer region of the pattern in Fig. 1 that they no longer combine with .m. in a statistically sound manner to point group *2mm*. More results and discussions of the point symmetry analysis of the electron diffraction pattern in Fig. 1 are provided in [12].

### III. POSSIBLE APPLICATIONS OF THE NEW METHOD

Due to its high sensitivity, the information-theoretic electron diffraction pattern based classification/quantification method for 2D point symmetries is very useful for the precise alignment of low indexed zone axes parallel to the optical axis of the TEM. A binary type method for distinguishing between quasicrystals and their rational approximants on the basis of the amplitude maps of the discrete Fourier transforms of a high-resolution image from a crystal or quasicrystal is mentioned in [13] and the expanded on line (arXiv) version of [8]. The same tasks can obviously be accomplished by the application of an information-theoretic classification and quantification method that works with electron diffraction spot and non-overlapping, essentially featureless, disk patterns. Beyond obvious applications in electron crystallography [10], a novel contrast mechanism for 2D scanning TEM on a 2D grid [14] would benefit from the incorporation of the new method, as discussed in [12]. That contrast mode has been referred to as "symmetry 4D STEM".

### IV. SUMMARY AND CONCLUSIONS

We quantified the evidence that the diffraction pattern in Fig. 1 originated from a crystal with a rectangular-centered Bravais lattice (in spite of the extracted pseudo-hexagonal lattice parameters). In addition, we classified the central part of this diffraction pattern out to Abbe resolution 1.25 Å as featuring point symmetry *2mm*, allowing for an averaging over sets of four diffraction spots as they would have the same intensity in the absence of generalized noise.

The outer region of the diffraction pattern in Fig. 1 beyond the Abbe resolution of 1.25 Å features only point symmetry group .m.. There are, thus, only mirror-related pairs of highest resolution diffraction spots for which an averaging of their intensities makes sense. This is to a large extent due to missing spots around the margins of Fig. 1. Friedel pair spots are no longer classified as symmetry equivalent in the outer region of this figure.

Because point symmetries in diffraction patterns can now be objectively quantified, i.e. their amount measured and values for the measurement uncertainty given, there will be many applications of the new method in years to come. A generalization of the presented 2D information-theoretic method to diffraction data from crystals in three dimensions would be very useful since the traditional symmetry classifications in the fields of single-crystal X-ray and neutron crystallography contain elements of subjectivity as well.